\documentclass[pre,amsmath,amssymb,twocolumn,floatfix,showkeys,groupedaddress]{revtex4}

\usepackage{graphicx}
\usepackage{hyperref}

\renewcommand{\vec}[1]{\ensuremath{\mathbf{#1}}}
\renewcommand{\eqref}[1]{(eq:\ref{#1})}

\renewcommand{\[}{\begin{equation}}
\renewcommand{\]}{\end{equation}}
\newcommand{\comment}[1]{}

\begin{document}

%
%
%
%
%
%
%
%
%
%

\title{Crystallography on Curved Surfaces}
\author{Vincenzo Vitelli}
\email[E-mail: ]{vitelli@sas.upenn.edu}
\altaffiliation[Permanent Address: ]{Department of Physics and Astronomy, U Pennsylvania\\
   209 South 33rd Street, Philadelphia, PA 19104-6396}
\author{Julius B. Lucks}
\author{David R. Nelson}
\affiliation{Lyman Laboratory of Physics, Harvard University, Cambridge Massachusetts 02138}



\begin{abstract}
We study static and dynamical properties that distinguish two dimensional crystals constrained to lie 
on a curved substrate from their flat space counterparts. A generic mechanism of dislocation unbinding in the presence of varying Gaussian curvature is presented in the context of a model surface amenable to full  analytical treatment. We find that glide diffusion of isolated dislocations is suppressed by a binding potential of purely geometrical origin. Finally, the energetics and biased diffusion dynamics of point defects such as vacancies and interstitials is explained in terms of their geometric potential.
\end{abstract}

\maketitle


\section{Introduction} 

                             
The physics of two dimensional crystals on curved substrates is emerging as an intriguing route to the engineering of self assembled systems such as the ``colloidosome'', a colloidal armor used for drug delivery \cite{baus03}, or devices based on ordered arrays of block copolymers which are a promising tool for ``soft lithography''  \cite{Park-Chaikin97,Kramer05}. 
Curved crystalline order also affects the mechanical properties of biological structures like clathrin-coated pits \cite{Heus89, kohy03, schn05} or 
HIV viral capsids \cite{Ganser99,Li2000} whose irregular shapes appear to induce a non-uniform distribution of disclinations in their shell \cite{Toan05}.

In this article, we present a theoretical and numerical study of point-like defects in a ``soft'' crystalline monolayer grown on a rigid substrate of $varying$ Gaussian curvature. We suppose that the monolayer has lattice constant of order, say,  10 nm or more. The substrate can then be assumed smooth, as would be the case for monolayers composed of di-block copolymers \cite{Park-Chaikin97,Kramer05}. Disclinations and dislocations are important topological defects that induce long range disruptions of orientational or translational order respectively \cite{Nelsbook,Chaikin-Lubensky-book}. Disclinations are points of local 5- and 7-fold symmetry in a triangular lattice (labeled by topological charges $q=\pm\frac{2\pi}{6}$ respectively), while dislocations are disclination dipoles characterized by a Burger's vector, $\vec{b}$, defined as the amount by which a circuit drawn around the dislocation fails to close (see Figure \ref{fig:dislocpotential} inset). Other point defects such as vacancies, interstitials or impurity atoms create shorter range disturbances that introduce only local stretching or compression in the lattice (see Figure \ref{fig:interstitials}). Such defects are important for particle diffusion and relaxation of concentration fluctuations.

These particle-like objects interact not only with each other, but also with the curvature of the substrate via a one-body geometric potential that depends on the particular type of defect \cite{sach84,bowi2000}. These geometric potentials are in general $non-local$ functions of the Gaussian curvature that we determine explicitly here for a model surface shaped as a ``Gaussian bump''. An isolated bump of this kind models long wavelength undulations of a lithographic substrate, has regions of both positive and negative curvature, and yet is simple enough to allow straightforward analytic and numerical calculations. The presence of these geometric potentials triggers defect unbinding instabilities in the ground state of the curved space crystal, even if no topological constraints on the net number of defects exist. Geometric potentials also control the dynamics of isolated dislocations by suppressing motion in the glide direction. Similar mechanisms influence the equilibrium distribution and dynamics of vacancies, interstitials and impurity atoms.


\section{Basic Formalism \label{basic}} 


The in-plane elastic free energy of a crystal embedded in a gently curved frozen substrate, given in the Monge form by $\vec{r}(x,y)=(x,y,h(x,y))$, can be expressed in terms of the Lam\'{e} coefficients $\mu$ and $\lambda$ \cite{Landau-book}
\begin{equation}
 F=\frac{1}{2} \int\! dA \!\!\!\! \quad \sigma_{ij}(\vec{x}) u_{ij}(\vec{x}) =  \int\! dA \!\!\!\!\! \quad \left( \mu \!\!\!\!\! \quad u_{ij} ^{2}(\vec{x}) + \frac{\lambda}{2} \!\!\!\!\! \quad u_{kk} ^2(\vec{x})   \right) \!\!\!\! \quad ,
\label{eq:lame}
\end{equation}
where $\vec{x}=\{x,y\}$ represents a set of standard cartesian coordinates in the plane and $dA=dx dy \sqrt{g}$, where $\sqrt{g}=\sqrt{1+(\partial_{x}h)^2 + (\partial_{y}h)^2}$. In Eq.(\ref{eq:lame}) the stress tensor $\sigma _{ij}(\vec{x})=2\mu u_{ij}(\vec{x})+ \lambda \delta_{ij} u_{kk}(\vec{x})$ was written in terms of the strain tensor $u_{ij} (\vec{x})=\frac{1}{2}\left[\partial_{i} u_j (\vec{x})+ \partial_{j} u_i (\vec{x})+ A_{ij} (\vec{x})\right]$. The latter contains an additional term $A_{ij}(\vec{x})=\partial_{i}h(\vec{x}) \partial_{j}h(\vec{x})$ (compared to its flat space counterpart) that couples the gradient of  the displacement field $u_{i}(\vec{x})$ to the geometry of the substrate as embedded in the gradient of the substrate height function $h(\vec{x})$. We will often illustrate our results for a single Gaussian bump  whose height function is given by 
\begin{equation}
h(\vec{x})= \alpha \!\!\!\!\! \quad x_{0} e^{-\frac{r^2}{2}}  \!\!\!\! \quad ,
\label{eq:constraint}
\end{equation}
where $r \equiv \frac{|\vec{x}|}{x_{0}}$ is the dimensionless radial coordinate \cite{ViteNels04}.  The aspect ratio $\alpha$ provides a dimensionless control parameter to measure deviations from flatness. 

In Eq.(\ref{eq:lame}) and the rest of this paper, we adopt the ordinary flat space metric (e.g. setting $dA\approx dx dy$) and absorb all the complications associated with the curved substrate into the tensor field $A_{ij}(\vec{x})$ that resembles the more familiar vector potential of electromagnetism. Like its electromagnetic analog, the curl of the tensor field $A_{ij}(\vec{x})$ has a clear physical meaning and is equal to the Gaussian curvature of the surface $G(\vec{x}) = -\epsilon_{il}\epsilon_{jk}\partial_{l}\partial_{k}A_{ij}(\vec{x})$ where $\epsilon_{ij}$ is the antisymmetric unit tensor ($\epsilon_{xy}=-\epsilon_{yx}=1$) \cite{sach84}. The consistency of our perturbative formalism to leading order in $\alpha$ is demonstrated in Appendix \ref{app-Monge} (Supplemental Materials).   
  
Minimization of the free energy in Eq.(\ref{eq:lame}) with respect to the displacements $u_i(\vec{x})$  naturally leads to the force-balance equation, $\partial _{i} \sigma_{ij}(\vec{x})=0$.  If we write $\sigma_{ij}(\vec{x})$ in terms of the Airy stress function $\chi(\vec{x})$
\begin{equation}
\sigma_{ij} (\vec{x})=\epsilon_{il}\epsilon_{jk}\partial_{l}\partial_{k} \chi(\vec{x})
\label{eq:chi1}
\end{equation}
then the force balance equation is automatically satisfied, 
\begin{equation}
\label{eq:force-bal}
\partial _{i} \sigma_{ij}=\epsilon_{jk}\partial_k [\partial_1,\partial_2]\chi(\vec{x}) = 0 \!\!\!\! \quad ,
\end{equation}
since the commutator of partial derivatives is zero. However, we must be able to extract from $\chi(\vec{x})$ the correct $u_{ij}(\vec{x})$ that incorporates the geometric constraint of Eq.(\ref{eq:constraint}) and accounts for the presence of any defects. These requirements are enforced by solving a bi-harmonic equation for $\chi(\vec{x})$ whose source is controlled by the varying Gaussian curvature of the surface $G(\vec{x})$  and the defect density $S(\vec{x})$ \cite{Nelsbook,sach84} 
\begin{equation}
\frac{1}{Y} \Delta^{2} \chi (\vec{x})= S(\vec{x})-G(\vec{x}) \!\!\!\! \quad .
\label{eq:vonKarman}
\end{equation}
In the special case $S(\vec{x})=0$, we denote the solution of Eq.(\ref{eq:vonKarman}) as $\chi^{G}(\vec{x})$ where the superscript $G$ indicates that the Airy function and the corresponding stress tensor $\sigma^{G}_{ij}(\vec{x})$ describe elastic deformations caused by the Gaussian curvature $G(\vec{x})$ only, without contribution from the defects. The stress of geometric frustration, $\sigma^{G}_{ij}(\vec{x})$, is a $non-local$ function of the curvature of the substrate and plays a central
role in our treatment of curved space crystallography. The Young modulus $Y=\frac{4\mu(\mu + \lambda)}{2 \mu + \lambda}$  \cite{Nelsbook,Chaikin-Lubensky-book} naturally arises upon recasting (up to boundary terms) the free energy in Eq.(\ref{eq:lame}) as a simple functional of the scalar field $\chi(\vec{x})$ 
\begin{equation}
 F=\frac{1}{2 Y} \int\! dA \!\!\!\!\! \quad \left(\Delta \chi(\vec{x}) \right)^2 \!\!\!\! \quad .
\label{eq:chi}
\end{equation}

The source $S(\vec{x})$ for a distribution of $N$ unbound disclinations with ``topological charges'' $\{q_{\alpha}=\pm \frac{2 \pi}{6}\}$ and $M$ dislocations with Burger's vectors $\{\vec{b}^{\beta}\}$ reads \cite{Nelsbook}
\begin{equation}
S(\vec{x}) = \sum_{\alpha=1}^{N} q_{\alpha} \delta (\vec{x},\vec{x} ^{\alpha}) + \sum_{\beta=1}^{M} \epsilon_{ij} b^{\beta} _{i} \partial _{j} \delta (\vec{x},\vec{x} ^{\beta}) \!\!\!\! \quad ,
\label{eq:s1}
\end{equation}
where $|\vec{b}|$ is equal to the lattice constant $a$ for dislocations with the smallest Burger's vector. For $N$ isotropic vacancies, interstitials or impurities, we have \cite{Nelsbook}
\begin{equation}
S(\vec{x}) = \frac{1}{2} \sum_{\alpha=1}^{N} \Omega_{\alpha} \!\!\!\!\! \quad \Delta \delta (\vec{x},\vec{x}^{\alpha})  \!\!\!\! \quad , \label{eq:s3}
\end{equation}
where $\Omega_{\alpha} \sim a^2$ is the local area change caused by including the point defect at position $\vec{x}^{\alpha}$. 

It is convenient to introduce an auxiliary function $V(\vec{x})$ that satisfies the Poisson equation $\Delta V(\vec{x}) = G(\vec{x})$ and vanishes at infinity where the surface flattens out.
In order to determine the geometric potential, $\zeta({\vec{x}^{\alpha}})$,  of a defect at $\vec{x^{\alpha}}$ 
we integrate by parts twice in Eq.(\ref{eq:chi}) and use Eq.(\ref{eq:vonKarman})  
to obtain $\Delta^2 \chi (\vec{x})$ and $\chi(\vec{x})$ in terms of the Green's function of the biharmonic operator. The geometric potential (on a deformed plane flat at infinity) follows from integrating by parts the cross terms involving the source and the Gaussian curvature with the result
\begin{eqnarray}
\zeta (\vec{x}^{\alpha}) = -Y \int \! dA' S(\vec{x}') \!\!\!\! \quad \int \! dA \frac{1}{\Delta_{\vec{x}\vec{x}'}} \!\!\!\! \quad V(\vec{x}) \!\!\!\! \quad . 
\label{eq:intermediate-step}
\end{eqnarray} 
This formula ignores defect self-energies \footnote{For dislocations, vacancies and interstitials the position-dependent self energies can be ignored compared to $\zeta (\vec{x}^{\alpha})$ because they are proportional to higher powers of the lattice spacing $a$.} and needs to be supplemented by boundary corrections, as discussed in Appendix \ref{app-boundary} (Supplemental Materials). 

In the following sections, we explicitly show that our results derived from Eq.(\ref{eq:intermediate-step}) can also be obtained by means of more heuristic arguments. According to this point of view, defects introduced in the curved space crystal are local probes of the preexisting stress of geometric frustration $\sigma^{G}_{ij}(\vec{x})$ to which they are coupled by intuitive physical mechanisms such as Peach-Koehler forces. 


\section{\label{frustration} Geometric Frustration} 


We start by calculating the energy of a relaxed defect-free two dimensional crystal on a quenched topography. In analogy with the bending of thin plates we expect some stretching to arise as an unavoidable consequence of the geometric constraints associated with the Gaussian curvature \cite{Landau-book}. The resulting energy of geometric frustration, $F_{0}$, can be estimated with the aid of Eq.(\ref{eq:vonKarman}) which, when $S(\vec{x})=0$, reduces to a Poisson equation whose source is given by $V(\vec{x})$
\begin{equation}
\frac{1}{Y} \Delta \chi ^{G}(\vec{x})= -V(\vec{x}) + H_{R}(\vec{x}) \!\!\!\! \quad .
\label{eq:vonKarman2}
\end{equation}   
where $H_{R}(\vec{x})$ is an harmonic function of $\vec{x}$ parameterized by the radius of the circular boundary $R$. As discussed in Appendix \ref{app-boundary}, $H_{R}(\vec{x})$ vanishes in the limit $R \gg x_{0}$ if free boundary conditions are chosen. Upon using the general definition of the Airy function (Eq. \ref{eq:chi1}), we obtain
\begin{equation}
\label{eq:sigma}
\sigma^{G}_{kk} (\vec{x})= \Delta \chi^{G}(\vec{x}) = -YV(\vec{x}) \!\!\!\! \quad .
\end{equation}
For a surface with azimuthal symmetry, like the bump, the only non vanishing components of 
the stress tensor of geometric frustration $\sigma^{G}_{ij}(\vec{x})$ read
\begin{equation}
\sigma ^{G}_{\phi \phi}(r) =  \frac{1}{x_{0}^2} \frac{\partial ^2  \chi ^{G}}{\partial r ^2 } \!\!\!\! \quad .
\label{eq:sigma_phiphi}
\end{equation}   
\begin{equation}
\sigma ^{G}_{rr}(r) =  \frac{1}{x_{0}^2 \!\!\!\!\! \quad r} \frac{\partial \chi ^{G}}{\partial r} \!\!\!\! \quad .
\label{eq:sigma_rr}
\end{equation}   
and Poisson's equation can be readily solved upon applying Gauss' theorem with the Gaussian curvature $G(r) \approx \frac{\alpha ^2 e^{- r^2}  (1-r^2)}{x_{0}^2}$ as a source \cite{ViteNels04}:  
\[
   V(r) = - x_{0}^2 \int_{r}^{\infty} \frac{dr'}{r'} 
                   \int_{0}^{r'} dr'' r'' G(r'')=-\frac{1}{4} \alpha ^{2} e^{-r^2}
 \!\!\!\! \quad . \label{eq:V}
\]
The extra factors of $x_{0}$ in the previous equations arise from expressing our results in terms of the dimensionless radial distance $r$.

Upon substituting Eq.(\ref{eq:vonKarman2}) in Eq.(\ref{eq:chi}), we have
\[
   F_{0} \simeq \frac{Y}{2} \int \! dA \!\!\!\! \quad
   V(r)^2 = \frac{Y \pi x_{0}^2 \alpha ^4 }{64}
   \!\!\!\! \quad . \label{eq:F0}
\]
The result in Eq.(\ref{eq:F0}) is valid to leading order in $\alpha$, consistent with the assumptions of our formalism after taking the limit $R \gg x_{0}$ (see Appendix B for finite size corrections in small systems). For an harmonic lattice, $Y=\frac{2}{\sqrt{3}}k$, where $k$ is an effective spring constant that can be extracted from more realistic inter-particle potentials \cite{seun88}. For colloidal particles, $k$ is typically two orders of magnitude larger than $k_{B} T /a^2$, where $T$ is room temperature \cite{Lipo05}. Our numerical calculations of $F_{0}$ in fixed-connectivity harmonic solids are in good agreement with the small $\alpha$ expansion in Eq.(\ref{eq:F0}) as long as the aspect ratio $\alpha$ is around $1/2$ or lower (see Appendix \ref{app-Monge} for a numerical plot of $F_{0}$ versus $\alpha$).  An immediate implication of the geometric frustration embodied in Eq. (\ref{eq:V}) and (\ref{eq:F0}) is that nucleation of crystal domains on the bump will take place preferentially away from the top in regions where the surface flattens out.

%
\section{Geometric potential for dislocations \label{sec:dislocpotential}}
%

\begin{figure}
\includegraphics{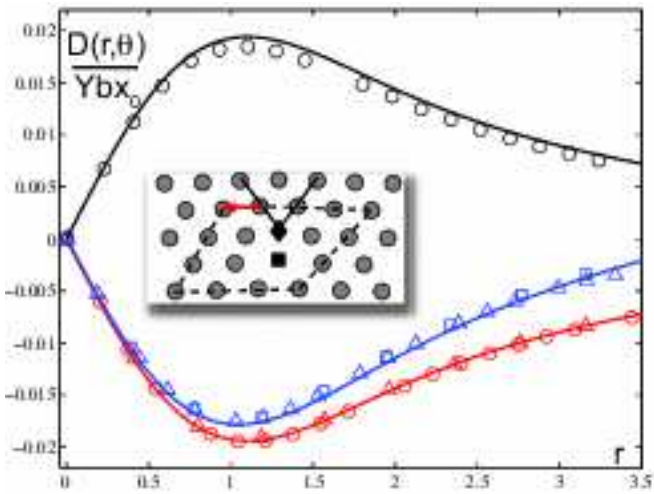}
\caption{\label{fig:dislocpotential} The dislocation potential $D(r,\theta=\pm \frac{\pi}{2})$ in Eq.(\ref{eq:D1}) (including boundary corrections) is plotted as a continuous line for a Gaussian bump parameterized by $\alpha = 0.5$ in the limit $R >> x_0 >> a$.  Open symbols represent the numerical minimization of a fixed connectivity harmonic model for which the separation of the +/- disclination pair representing the dislocation is fixed while varying the dimensionless  radial distance $\frac{|\vec{x}|}{x_{0}}$ of the dislocation from the bump. An angle $\theta = \frac{\pi}{2}$ indicates that the + disclination is closer to the top of the bump. Lower branch: $\theta=\pi/2$, $R/x_0=$4 (blue), 8 (red); and $x_0/a=$ 10 ($\bigcirc$), 20 ($\triangle$), 40 ($\Box$). Upper branch: $\theta=-\pi/2$, $x_0/R = 8$, $x_0/a=10$.  Inset: A schematic view of a dislocation with filled symbols representing six- (circles), five- (diamond) and seven- (square) coordinate particles.  Also depicted are the two rows of extra atoms emanating from the five-coordinated particle.  The Burger's vector is shown as a red arrow, completing the circuit around the dislocation (dashed line).}
\end{figure}

The energy of a two dimensional curved crystal $with$ defects will include the frustration energy, the inter-defect interactions (to leading order these are unchanged from their flat space form, see Appendix \ref{app-Monge}), possible core energies and a characteristic, one-body potential of purely geometrical origin that describes the coupling of the defects to the curvature given by Eq.(\ref{eq:intermediate-step}).  
The geometric potential of an isolated dislocation, $\zeta (\vec{x}) \equiv D(\vec{x},\theta)$, is a function of its position as well as of the angle $\theta$ that the Burger's vector $\vec{b}$ forms with respect to the radial direction (in the tangent plane of the surface. See Figure \ref{fig:dislocpotential}.) Upon setting all $q_{\alpha}=0$ in Eq.(\ref{eq:s1}) and substituting into Eq.(\ref{eq:intermediate-step}), we obtain, for an isolated dislocation,  the resulting function $D(r=\frac{|\vec{x}|}{x_{0}},\theta)$
\begin{eqnarray}
   D(r,\theta) &=& - Y b_{i} \!\!\!\!\! \quad \epsilon_{ij} \!\!\!\!\! \quad \partial_{j} \int \! dA' \!\!\!\! \quad \frac{1}{\Delta_{\vec{x}\vec{x'}}} \!\!\!\! \quad
  \left(V(\vec{x'})+\frac{\alpha ^2 x_{0}^2}{4 R^2}\right)  \nonumber\\ & \approx & Y \!\!\!\!\! \quad  b \!\!\!\!\! \quad x_{0} \!\!\!\!\! \quad \frac      
         {\alpha^{2}}{8}  \sin{\theta} \left[ \left( \frac{e^{-r^{2}} - 1}{r}\right) + \left(\frac{x_{0}}{R}\right)^2 r \right] \!\!\!\! \quad . \label{eq:D1}
\end{eqnarray}
In view of the azimuthal symmetry of the surface, Gauss' theorem as expressed in Eq.(\ref{eq:V}), was employed in  deriving the second equality in Eq.(\ref{eq:D1}) which is a function only of the dimensionless radial coordinate $r$. The first term in Eq.(\ref{eq:D1}) corresponds to the $R \rightarrow \infty$ geometric potential obtained from Eq.(\ref{eq:intermediate-step}), while the second term is a finite size correction arising from a circular boundary of radius $R$ (see Appendix \ref{app-boundary} for a detailed derivation). Eq.(\ref{eq:D1}) is valid to leading order in perturbation theory, consistent with the small $\alpha$ approximation adopted in this work (see Appendix \ref{app-Monge}).  

In Fig. \ref{fig:dislocpotential} we present a detailed comparison between the theoretical predictions for the geometric potential $\frac {D(r,\theta)}{Y b x_{0}}$ plotted versus $r=\frac{|\vec{x}|}{x_{0}}$ as 
continuous lines and numerical data from constrained minimization of an harmonic solid on a bump with $\alpha=0.5$, under conditions such that $R \gg x_{0} \gg a$. (See Appendix \ref{app-numerics} for a discussion of our numerical approach).
The lower and upper branches of the graph are obtained from Eq. (\ref{eq:D1}) by setting $\theta=\pm \frac{\pi}{2}$ and letting $\frac{R}{x_{0}}$ equal to $4$ (blue curve) or $8$ (red curve). Indeed, the (scaled) data from simulations with different choices of $\frac{x_{0}}{a}$ (see caption) collapse on the two master-curves according to their ratio of $\frac{R}{x_{0}}$. Two different curves arise because the dislocation interacts with the curvature directly and via its image. The image-mediated interaction is given by the $R$ dependent term in Eq.(\ref{eq:D1}).
The $\sin \theta$ dependence of $D(r,\theta)$ on the direction of the Burgers vector is revealed by Fig. \ref{fig:dislocpotential}, since the upper branch of the graph corresponding to the unstable equilibrium $\theta=-\frac{\pi}{2}$ is approximately symmetric to the lower one corresponding to $\theta=\frac{\pi}{2}$. 

The analogy between the geometrical potential of the dislocation and the more familiar interaction of an electric dipole in an external field can be elucidated by regarding the dislocation as a charge neutral pair of disclinations whose dipole moment $qd_{i}=\epsilon_{ij} b_{j}$ is a lattice vector perpendicular to $\vec{b}$ that connects the two points of 5 and 7-fold symmetry. The geometric potential, $U(r)$, of a disclination of topological charge $q$ interacting with the Gaussian curvature satisfies the Poisson equation $\Delta U(r) = - q V(r)$ as can be seen by substituting the source in Eq.(\ref{eq:s1}) with all $\vec{b}^{\beta}=0$ into Eq.(\ref{eq:intermediate-step}) . 
For small $r$, positive (negative) disclinations are attracted (repelled) from the center of the bump by the integrated background source $V(r)$ which increases for $r \le 1$ like $\alpha ^2 r^2$ and is multiplied by the $2D$ electric field $\frac{1}{r}$, resulting in a geometric force that increases linearly in $r$. If the positive disclination within the dipole is closer to the top,  the force it experiences will be opposite and slightly less than the one experienced by the negative disclination that is further away from the top. As a result a ``tidal'' force will push the dislocation (as a whole) down hill as shown by the lower branch of the plot of Fig. \ref{fig:dislocpotential}. For large $r$, however, the source $V(r)$ saturates and the attractive force exerted on the positive disclination  wins and drags the dislocation towards the bump. The minimum of the geometric potential occurs at $|\vec{x}_{min}|\approx1.1 \!\!\!\! \quad x_{0}$, close to the circle of zero Gaussian curvature $|\vec{x}|=x_{0}$, as a result of the competing interactions of the two disclinations comprising the dislocation. If the orientation of the disclination dipole is flipped, the geometric potential changes sign. Similarly, if the sign of the Gaussian curvature is reversed, the sign of the geometric interaction flips. As a result, a dislocation close to a saddle will have its Burger vector oriented so that the closest disclination to the saddle is 7-fold coordinated.
\begin{figure}[h]
		\begin{center}
		\includegraphics[scale=1]{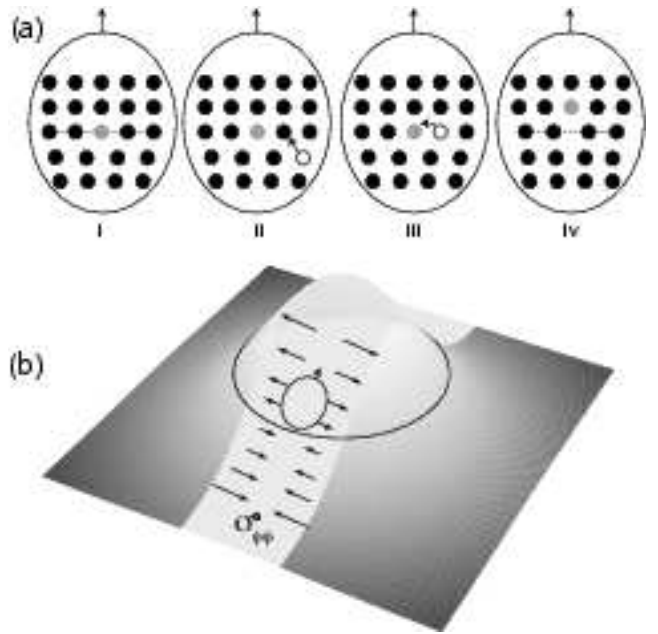}
	\end{center}
	\caption{Schematic of dislocation climb for the case of a square lattice. (a) Vacancy-mediated dislocation climb depicted in four steps: (i) Initial configuration consisting of an isolated dislocation (grey) (bonds in the third row are highlighted with dashed lines); (ii-iii) Motion of a vacancy (white) to the dislocation; (iv) Motion of the dislocation in the upward radial direction to create a configuration analogous to (i). Comparison of the third row of atoms in panel (i) and (iv) reveals that this process results in over-stretching of the bonds by a length of order of the lattice spacing $a$.  (b) On a bump (outline of patch shown), the process represented in (a) is energetically unfavorable close to the top of the bump because there is a tensile stress $\sigma ^{G} _{\phi \phi}$ (force per unit length indicated by arrows) that hinders further stretching of the bonds and tries to push the dislocation downward with a Peach-Koehler force $\sigma ^{G} _{\phi \phi}$$b$ where the Burger's vector $|\vec{b}| \approx a$. The force changes sign at a distance of approximately $x_{0}$ from the top of the bump (black ring) where $\sigma ^{G} _{\phi \phi}$ switches sign or if the direction of $\vec{b}$ is reversed.}
	\label{fig:disloc_climb}
\end{figure}

The physical origin of the dislocation potential can be understood heuristically by considering dislocation climb, or the motion of the dislocation in the radial direction (Figure \ref{fig:disloc_climb}). According to standard elasticity theory, a dislocation in an external stress field $\sigma_{ij}(\vec{x})$ experiences a Peach-Koehler force, $\vec{f}(\vec{x})$, given by $f_{k}(\vec{x})=\epsilon_{kj}b_{i} \sigma_{ij}(\vec{x})$ \cite{Kleman-book,Weertman-book}. Similarly, a dislocation introduced into the curved 2D crystal will experience a Peach-Koehler force as a result of the pre-existing stress field of geometric frustration $\sigma^{G}_{ij}(\vec{x^{\alpha}})$ whose non-diagonal components vanish. This interpretation is consistent with the geometric potential derived in Eq.(\ref{eq:D1}), provided  we use Eq. (\ref{eq:vonKarman2}) to write $D(\vec{x}) = b_i\epsilon_{ij}\partial_j\chi^{G}$.  With $\vec{b}$ along its minimum orientation (azimuthal counter-clockwise), we obtain a radial Peach-Koehler force of magnitude $f(r)=-b \!\!\!\!\! \quad \sigma^{G}_{\phi \phi}(r)$ that matches $- \frac{1}{x_0}\frac{\partial D(r)}{\partial r}=- \!\!\!\!\! \quad \frac{b}{x_0^2}\frac{\partial ^2 \chi^{G}(r)}{\partial r^2}$ (see Eq.(\ref{eq:sigma_phiphi})). 

%
\section{Dislocation Unbinding\label{sec:disloc_unbinding}}
%

If the $2D$ crystal is grown on a substrate which is sufficiently deformed, the resulting elastic strain can be partially relaxed  by introducing unbound dislocations into the ground state \cite{sach84,ViteNels04}.   Here we present a simple estimate of the threshold aspect ratio, $\alpha_{c}$, necessary to  trigger this instability. Boundary effects will be ignored in what follows
by letting $R \rightarrow \infty$ in Eq.(\ref{eq:D1}). 

\begin{figure*}
\includegraphics{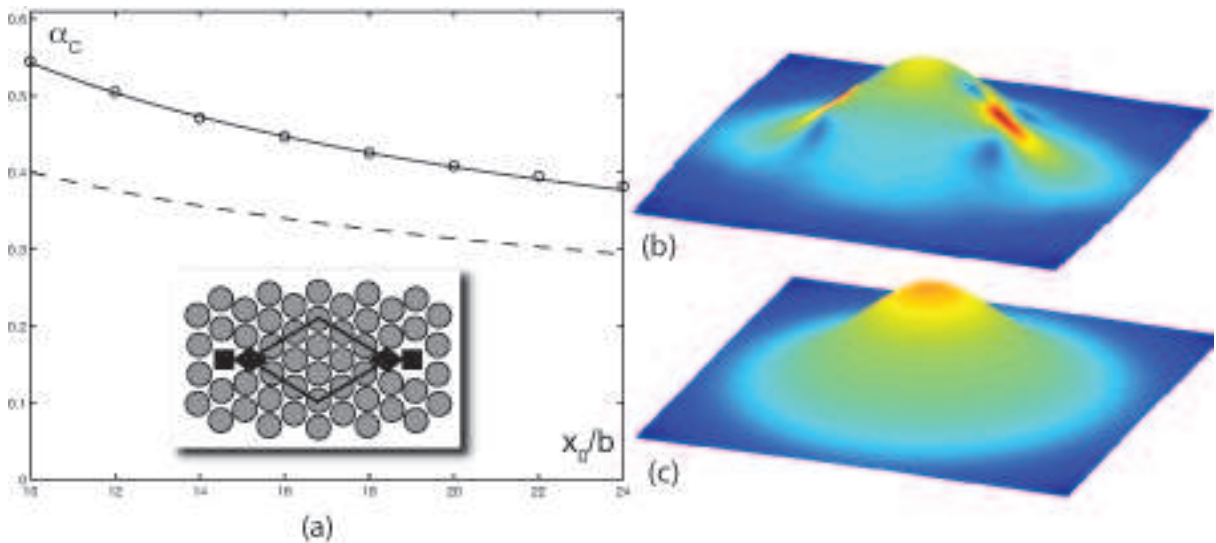}
\caption{\label{fig:dislocation_unbinding_strain} (a) Dislocation unbinding critical aspect ratio, $\alpha_c$, as a function of $\frac{x_{0}}{b}$.  The theoretical estimate (\ref{eq:dislocation_unbinding}) is plotted vs. $x_0$ for core energies $E_c = 0$ (dashed) and the best fit $E_c = 0.1Y b^2$ (solid).  Numerical values (circles) were obtained as described in Section \ref{sec:disloc_unbinding} and Appendix \ref{app-numerics}. Inset: particle configuration projected in the plane for a dislocation pair straddling the bump.  The extra atoms associated with the dislocations are highlighted with black lines.  (b) Log of the numerical strain energy density on a bump for the configuration shown in the inset. (c) The same quantity for a defect-free bump. Both plots were constructed numerically with $x_0 = 10$, $\alpha = 0.7 > \alpha_c$.  Red represents high strain.}
\end{figure*}

Consider two dislocations located at $\vec{x}_1$ and $\vec{x}_2$ a distance 
of approximately $x_{0}$ from the center of the bump on opposite sides (see the inset of Fig. \ref{fig:dislocation_unbinding_strain}). 
Their  disclination-dipole moments are opposite to each other and aligned in the radial direction so that  the two (antiparallel) Burger's vectors are perpendicular to the separation vector in the plane $\vec{x} _{12} \equiv \vec{x}_1 - \vec{x} _2$. In this case, the interaction between the dislocations reduces to $V_{12}\approx \frac{Y}{4 \pi} \!\!\!\! \quad b^2 \!\!\!\! \quad \ln \left(\frac{|\vec{x} _{12}|}{a} \right)$ \cite{Nelsbook,Chaikin-Lubensky-book}.
The instability occurs when the energy gain from placing each dislocation in the minima of the potential  $D(\vec{x})$ given by Eq.(\ref{eq:D1}) outweights the sum of the work needed to tear them apart plus the core energies $2 E_{c}$. The critical aspect ratio at the threshold which results is given by
\begin{eqnarray}
   \label{eq:dislocation_unbinding}
   \alpha_{c}^2 \approx c \!\!\!\! \quad \frac{b}{x_{0}} \!\!\!\! \quad \ln
   \left(\frac{x_{0}}{b \!\!\!\!\! \quad '} \right) \!\!\!\! \quad , \label{eq:alphac}
\end{eqnarray}
where $b' \equiv \frac{b}{2} \!\!\!\!\! \quad e^{-\frac{8 \pi E_{c}}{Yb^2}}$ and $c \approx 1/2$.  Note that the core energy $E_c$ is determined by the microscopic physics of the particular system under study.

In Fig. \ref{fig:dislocation_unbinding_strain}a we present a comparison between Eq. (\ref{eq:dislocation_unbinding}) and numerical results. For each value of $\frac{x_0}{b}$, the corresponding $\alpha_c$ is obtained numerically by comparing the energy of a lattice without defects to the configuration with the two dislocations in their equilibrium positions. 
This interpretation for the origin of the instability is corroborated by the (numerical) strain energy density plots of Fig. \ref{fig:dislocation_unbinding_strain}b and \ref{fig:dislocation_unbinding_strain}c , where it is shown that introducing the pair of dislocations reduces the strain energy density on top of the bump at the price of creating some large, but localized strains around the dislocation cores where $u_{ij}$ diverges. In the continuum limit $b \ll x_{0}$, very small deformations are enough to trigger the instability. This is the regime in which our perturbation treatment applies. 
As $\alpha$ is increased even further, a cascade of dislocation-unbinding transitions
occurs involving larger numbers of dislocations and more complicated equilibrium arrangements of zero net Burger vector. For sufficiently large aspect ratios, we expect that the dislocations tend to line up in grain boundary scars similar to the ones observed in spherical crystals \cite{baus03}. This scenario is consistent with preliminary results from Monte Carlo simulations in which the fixed-connectivity constraint is lifted and more complicated surface morphologies are considered \cite{Hex-Baltimore}.

%
\section{Glide suppression \label{glide}} 
%

\begin{figure*}
\includegraphics{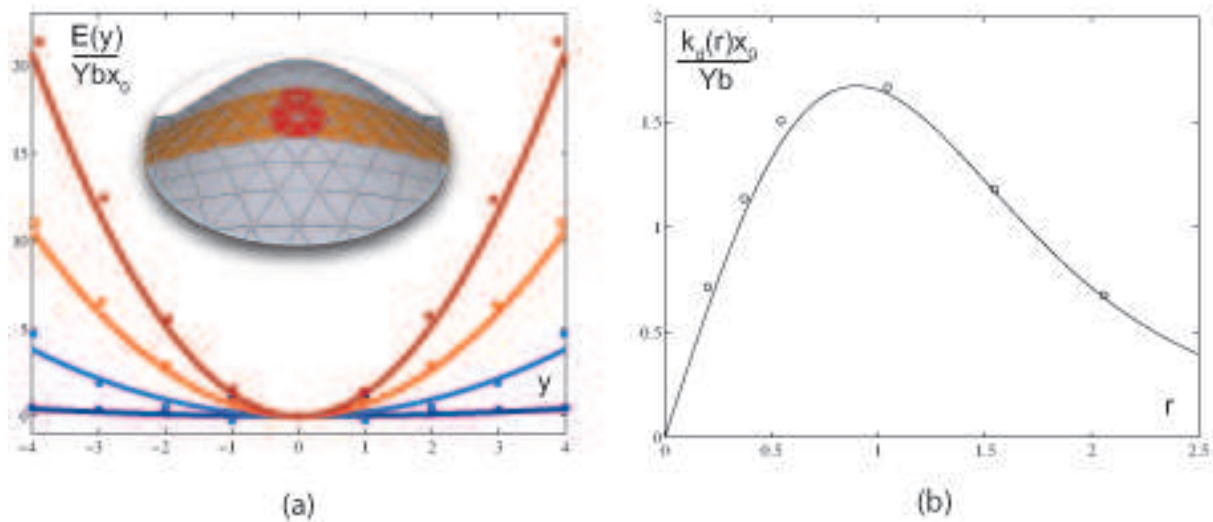}
\caption{\label{fig:glide} Dislocation glide.  (a) Filled circles represent numerical glide energies vs. distance along the glide direction y (units of a) for $x_0/a=10$, $R/x_0=8$, $r=0.5$ (at $y=0$): $\alpha=$ 0.1 (dark blue), 0.3 (blue), 0.5 (orange) and 0.7 (red).  Solid lines represent $E(y)=\frac{1}{2}k_{d}y^2$, with $k_d$ determined from Eq. (\ref{eq:k}).  The energy is scaled by $10^{-4}$.  Inset:  schematic of a dislocation (red) on a Gaussian bump.  The glide path is highlighted in orange. (b) The curvature-induced glide-suppression spring constant $k_d(r)$ (\ref{eq:k}) (solid line) is plotted versus scaled in-plane distance from the top of the bump $r = |\vec{x}|/x_0$ for $x_0=10 a, \alpha=0.5$.  Open symbols represent numerical results found by fitting similar data as in (a) to parabolic curves in the glide coordinate.  The ordinate axis is scaled by $10^{-2}$.}
\end{figure*}

The dynamics of dislocations proceeds by means of two distinct processes: glide and climb. Glide describes motion along the direction defined by the Burger's vector; in flat space glide requires a very low activation energy and is the dominant form of motion at low temperature (see Fig. \ref{fig:glide}a inset). 
Climb, or motion perpendicular to the Burger vector requires diffusion of vacancies and interstitials (Figure \ref{fig:disloc_climb}) and is usually frozen out relative to glide which involves only local rearrangements of atoms.
On a curved surface, the geometric potential $D(r, \theta)$ imposes constraints on the glide dynamics of isolated dislocations, in sharp contrast 
to flat space where only small energy barriers are present due to the periodic Peierls potential \cite{Kleman-book}. 

As the dislocation represented in the inset of Fig. \ref{fig:glide}a moves in the glide direction, it experiences a restoring potential generated by the variation of the (scaled) radial distance and the deviation from the radial alignment of the dislocation dipole. For a small transverse displacement, $y$, the harmonic potential $U(y)=\frac{1}{2}k_{d} y^2$ is controlled by a radial, position-dependent effective spring constant, $k_{d}$ which depends on the radial coordinate $r$. Upon expanding Eq.(\ref{eq:D1}) to leading order in $y$ and $\alpha$, we obtain 
\begin{eqnarray}
   k_{d}(r) \approx \frac{{\alpha}^{2} Y b}{4 \!\!\!\! \quad x_{0}} \left(\frac{1- (1+r^2)e^{-r^2}}{r^3}\right) \!\!\!\!    
   \quad .
   \label{eq:k}
\end{eqnarray} 
The harmonic potential $\frac{1}{2} k_{d} y^2$ is shown in Fig. \ref{fig:glide}a where the data obtained from numerical minimization of the harmonic lattice is explicitly compared to the prediction of Eq.(\ref{eq:k}) for different values of the aspect ratio. Note that the effective spring constant plotted in Fig. \ref{fig:glide}b vanishes in the limit 
$\frac{b}{x_{0}} \rightarrow 0$ but can still be important for small systems since $Yb^2$ can be of the order of hundreds of $k_{B}T\equiv \beta^{-1}$ (see section \ref{frustration}). 
The confining potential plotted in Fig.  \ref{fig:glide}a is similar to the one experienced by a dislocation bound to a disclination \cite{Brui82,Lipo05}.
 
The resulting thermal motion in the glide direction of dislocations in this binding harmonic potential can be modeled by an over-damped Langevin equation for the glide coordinate $y$, leading to $\langle {\Delta y}^2 \rangle  = \frac{1 - e^{-2 \mu k_{d} t}}{\beta k_{d}}$, where $\mu$ is the dislocation mobility \cite{Brui82,Lipo05}. In the case of a bump geometry, the effective spring constant $k_{d}$ can be evaluated using Eq.(\ref{eq:k}). We emphasize, however, that the glide suppression mechanism considered here is not caused by the interaction of the dislocation with other defects but purely by the geometric interaction with the curvature of the substrate.

%
\section{Vacancies, Interstitials and Impurities \label{sec:vacancies}} 
\begin{figure}
\includegraphics{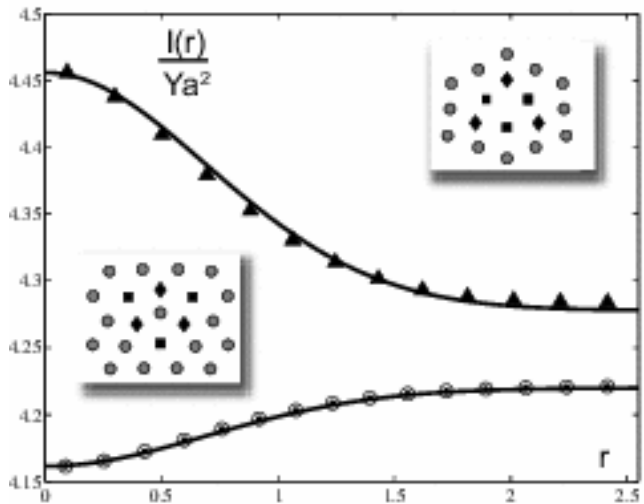}
\caption{\label{fig:interstitials} The  scaled potential energy ${I}$$(r)$/Y$a^2$ of interstitials (bottom) and vacancies (top) is plotted versus the scaled distance $r$ for the same bump as Fig. \ref{fig:dislocpotential}.  Solid lines are obtained  from Eq. (\ref{eq:I2}), where the area changes of interstitials and vacancies $\Omega^{\alpha}_i$ and $\Omega^{\alpha}_v$ were fit to the data.  Sample configurations are plotted in the insets,  where diamonds, circles and squares represent five, six and seven fold coordinated particles respectively. Filled and unfilled circles in the interstitial plot represent two distinct but energetically equivalent orientations of the disclination dipoles comprising the interstitial. They differ by swapping the 5 and 7-fold disclinations, which rotates the orientation of the filled, triangular plaquette by $\frac{\pi}{3}$.}
\end{figure}

%
We now turn to a derivation of the geometric potential, ${I}$$(\vec{x})$, for interstitials, isotropic vacancies and impurities. Inspection of  Fig. \ref{fig:interstitials} reveals that an interstitial (vacancy) can be viewed either as the product of locally adding (removing) an atom to the lattice or as a composite object made up of three disclination dipoles.
In order to derive ${I}$$(\vec{x})$, we substitute the source term of Eq.(\ref{eq:s3}) into Eq.(\ref{eq:intermediate-step}) and integrate by parts twice. The result reads
\begin{eqnarray}
   I(\vec{x}) \approx \frac{Y}{2} \Omega V(\vec{x}) \!\!\!\! \quad ,
   \label{eq:I2}
\end{eqnarray}
where boundary terms and a position independent nucleation energy have been dropped. The constant $\Omega$ represents the area excess or deficit associated with the defect. In Fig. \ref{fig:interstitials} a comparison of Eq.(\ref{eq:I2}) with the results obtained from mapping out ${I}$$(r)$ numerically is presented. The area changes $\Omega_{i}$ and $\Omega_{v}$ for interstitials and vacancies respectively were fit to the numerical data. We find that $\Omega_{v}$ is negative and greater in magnitude than $\Omega_{i}$. The large $r$ behavior of ${I}$$(r)$ indicates that the core energy of a vacancy in flat space is greater than the one of an interstitial for the harmonic lattice.
Interstitials tend to seek the top of the bump whereas vacancies are pushed into the flat space regions. From the definition of $V(r)$ in Eq.(\ref{eq:V}), we deduce that an interstitial (vacancy) is attracted (repelled) by regions of positive (negative) Gaussian curvature similar to the behavior of an electrostatic charge interacting with a background charge distribution given by $-G(r)$.  The function $V(r)$ controls the curvature-defect interaction for other types of defects such as
disclinations in liquid crystals and vortices in $^4$He films \cite{ViteNels04,VitePRL04}. The expression for $V(r)$ in Eq.(\ref{eq:V}) reveals that ${I}$$(r)$ is indeed a non local function of the Gaussian curvature determined as in electrostatics by the application of Gauss' law. Thus, the vacancy potential on a bump does not reach a minimum at the point where the Gaussian curvature is maximally negative but rather at infinity where the $integrated$ Gaussian curvature vanishes.

We now argue heuristically how a localized point defect can couple non-locally to the curvature and in the process we provide an alternative justification of Eq. (\ref{eq:I2}) analogous to the informal derivation of the dislocation potential. The energy cost of a defect at $\vec{x}^{\alpha}$, ${I}$$(\vec{x} ^{\alpha})$, due to local compression or stretching in the presence of an arbitrary elastic stress tensor  $\sigma_{ij}(\vec{x})$ is given by ${I}$$(\vec{x} ^{\alpha})= p (\vec{x} ^{\alpha}) \delta V$ where $\delta V$ is the local volume change and $p(\vec{x^{\alpha}})$ is the local pressure related to the stress tensor via $\sigma_{ij}(\vec{x} ^{\alpha})=-p(\vec{x} ^{\alpha}) \delta_{ij}$ for the case of an isotropic stress\cite{Landau-book}. In two dimensions, we have ${I}$$(\vec{x}^{\alpha})=-\sigma_{kk}(\vec{x}^{\alpha})\frac{\Omega ^{\alpha}}{2}$. We recover the result in Eq.(\ref{eq:I2}), by assuming that the local deformation $\Omega ^{\alpha}$ (induced by the nucleation of a point defect) couples to the preexisting stress of geometric frustration $\sigma^{G}_{kk}=-YV(\vec{x})$ (see Eq.(\ref{eq:sigma})), which is a non-local function of the Gaussian curvature. Elastic deformations created by the geometric constraint throughout the curved two dimensional solid are propagated to the position of the point defect by force chains spanning the \emph{entire} system. 
The point defect can then be viewed as a local probe of the stress field that does not measure the additional stresses induced by its own presence. 

Note that the geometrical potential of an isotropic point defect is unchanged if we swap the 5 and 7-fold disclinations comprising it (corresponding to a rotation of the point defect by 
$\frac{\pi}{3}$ around its center), as demonstrated for interstitials in the lower branch of Fig. \ref{fig:interstitials}. Contrast this situation with the clear dipolar character of the dislocation potential plotted in Fig. \ref{fig:dislocpotential}.  The more complicated case of non-isotropic point defects appears to still be captured qualitatively by Eq.(\ref{eq:I2}); this was explicitly checked by plotting the geometric potential of ``crushed vacancies'' \cite{jain2000}, which have both a lower symmetry and a lower energy than their isotropic counterparts. 

An arbitrary configuration of weakly interacting point defects will relax to its equilibrium distribution by diffusive motion in a force field  \textbf{f}(r)$=- \nabla {I}$$(r)$. In overdamped situations, this geometric force leads to a biased diffusion dynamics with drift velocity 
$v \sim \beta |f| a \frac{D}{a} \sim \frac{D Y \beta \Omega}{x_{0}}$, where $\beta =\frac{1}{k_{B}T}$ and $D$ is the defect diffusivity \cite{Kleman-book}.  Eventually,  a dilute gas of point defects equilibrates to a non-unform spatial density proportional to $e^{-\beta{{I}}(r)}$. 

\textit{Acknowledgments:} We are grateful to G. Chan, B. I. Halperin, A. Hexemer, Y. Kafri, E. Kramer and A. M. Turner for stimulating conversations. This work was supported by the National
Science Foundation, primarily through the Harvard Materials
Research Science and Engineering Laboratory via Grant No.
DMR-0213805 and through Grant No. DMR-0231631.
JBL acknowledges financial support from the Hertz Foundation.

\appendix

The supplemental materials are organized in the form of three Appendices A, B, and C.

\section{\label{app-Monge} Perturbation theory of curved crystals}
The starting point of our perturbative analysis of curved crystalline order is Eq.(\ref{eq:vonKarman}) which incorporates Gaussian curvature by adding an extra source to the defect term but it is nonetheless written using a flat space metric. To underscore the subtleties involved we write a covariant generalization of the force balance equation, $\partial _{i} \sigma_{ij}(\vec{x})=0$,  \cite{Heinbockel-book}
\begin{equation}
\frac{1}{\sqrt{g}}\frac{\partial}{\partial x^{i}}(\sqrt{g}\sigma^{i}_{j})-\Gamma ^{k} _{ji} \sigma ^{i} _{k}=0 \!\!\!\!\! \quad . \label{eq:balance-cov}
\end{equation}
where $g$ is the determinant of the metric tensor $g_{ij}$ and $\Gamma ^{k} _{ji}$ is the Christoffel symbol. Eq.(\ref{eq:balance-cov}) can be concisely written as 
$\sigma^{i} _{j ;i}=0$, where the semicolons indicate the covariant derivatives $D_{i}$ \cite{Heinbockel-book}. 

Strictly speaking, this equation cannot be solved by using the flat space trick of writing the stress tensor in terms of the Airy function because of a distinctive property of curved space: torsion \cite{Pete05,Davidreview}. An arbitrary vector $\vec{T}$ parallel transported around a closed loop on a surface of non vanishing Gaussian curvature is rotated from its original orientation \footnote{In fact, the net effect of parallel transport is equivalent to monitoring the change of $\vec{T}$ first in one direction then in the other followed by subtracting changes in the reverse order.}.  In differential form, this constraint reads 
\begin{equation}
\left[D_{i},D_{j}\right] T_{k}= G(\vec{x}) \!\!\!\!\! \quad \gamma_{ij} \!\!\!\!\! \quad \gamma_{k}^{m}  \!\!\!\!\! \quad T_{m} , \label{eq:commute}
\end{equation}
where $\gamma_{ij}=\frac{\epsilon_{ij}}{\sqrt{g}}$ denotes the antisymmetric tensor on a curved surface\cite{Davidreview}.  Upon substituting $T_{m}=\gamma_{m} \!\!\!\!\!\!\! \quad ^{n} \!\!\!\!\! \quad \partial_{n} \!\!\!\!\! \quad \chi(\vec{x})$ into Eq.(\ref{eq:commute}), we see that the covariant analogue of Eq.(\ref{eq:force-bal}) does not hold because the commutator of covariant derivatives (known as the torsion tensor) does not vanish.
\begin{figure}
\includegraphics{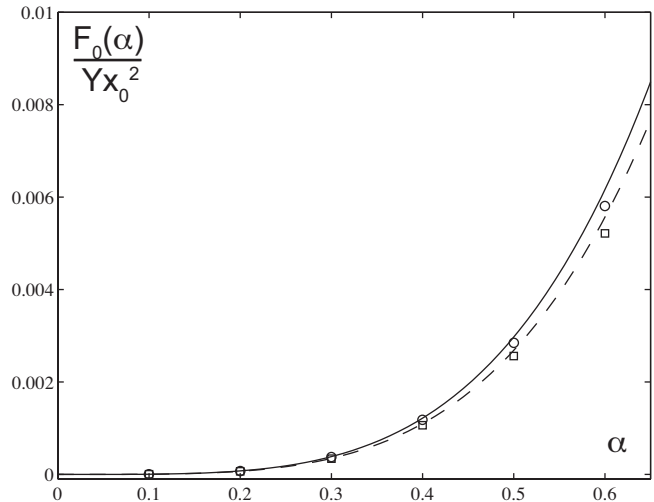}
\caption{\label{fig:F0} 
The geometric frustration energy $F_0(\alpha)$ (Eq. (\ref{eq:F0})) is plotted versus $\alpha$ for $x_{0}/R = 10/80$  ($\bigcirc$) and $10/40$ ($\Box$).  Solid ($x_{0}/R =10/80$) and dashed ($x_{0}/R=10/40$) lines indicate plots of Eq. (\ref{eq:F0}) using $Y = \frac{2}{\sqrt{3}}k$ \cite{seun88}, including the finite size corrections discussed in Appedix \ref{app-boundary} for these two conditions, respectively.}
\end{figure}

We can nonetheless make progress by noting first that the Gaussian curvature of a bump in Eq.(\ref{eq:commute}) is proportional to $\frac{\alpha^2}{x_{0}^2}$ and reads  \cite{ViteNels04}
\begin{equation}
G(r)=\frac{\alpha^2 e^{-r^2}}{x_{0}^2} \!\!\!\! \quad \frac{1-r^2}{\left(1+\alpha ^2 r^2 e^{-r^2}\right)^2}     \!\!\!\! \quad . 
\label{Gaussian_curvature}
\end{equation}
As a result, Eq. (\ref{eq:force-bal}) is in fact approximately correct for small $\alpha$. We first  consider the case $S(\vec{x})=0$, for which Eq.(\ref{eq:vonKarman}) reduces to one of the two celebrated Foppl-von Karman equations describing slightly deformed thin plates \cite{Landau-book}. For a frozen substrate, the second Foppl-von Karman equation (arising from the variation of the normal coordinate h(x,y)) determines the adhesion pressure  (normal to the surface) which is needed to constrain the $2D$ solid to the curved substrate \cite{Landau-book}. The Foppl-von Karman analysis rests on a consistent perturbation theory in $\alpha$ and predicts that $\chi(\vec{x})$ is ${O}$$(\alpha^2)$, as can be seen from Eq.(\ref{eq:vonKarman}) with $S(\vec{x})=0$. To this order, the commutator in Eq.(\ref{eq:commute}) indeed vanishes (the leading order corrections are 
${O}$$(\alpha^4)$) and one is justified in expressing $\sigma_{ij}(\vec{x})$ in terms of an Airy function $\chi(\vec{x})$. 

The situation is more delicate in the presence of defects because $\chi(\vec{x})$ no longer vanishes as $\alpha \rightarrow 0$ but tends instead to the flat space form (see Ref.\cite{Nelsbook,Chaikin-Lubensky-book} for explicit results). Dimensional analysis reveals that the commutator in Eq.(\ref{eq:commute}) is now proportional to $\frac{\alpha^2 Y}{x_{0}}\frac{b}{x_{0}}$ for dislocations and to $\frac{\alpha^2 Y}{x_{0}}\frac{\Omega}{x_{0}^2}$ for vacancies, interstitials and impurities where corrections of order $\ln\left(\frac{R}{a}\right)$ are ignored. 
Hence, the commutator in Eq.(\ref{eq:commute}), set to zero in the derivation of Eq.(\ref{eq:vonKarman}), appears to be of the same order in $\alpha$ as the curvature corrections to $\chi(\vec{x})$ that we wish to calculate. The commutator is still small, however, in the continuum limit $x_{0} \gg a$.  The use of Eq.(\ref{eq:vonKarman}) to study isolated $disclinations$ 
to leading order in $\alpha$ is more difficult to justify because in this case the commutator can be as large as  $\frac{\alpha^2 Y}{x_{0}}\frac{R}{x_{0}}$ . However, such an investigation on a surface with the topology of the plane is of limited interest, in view of the large energy cost of isolated disclinations (quadratic in R).
\begin{figure}
\includegraphics{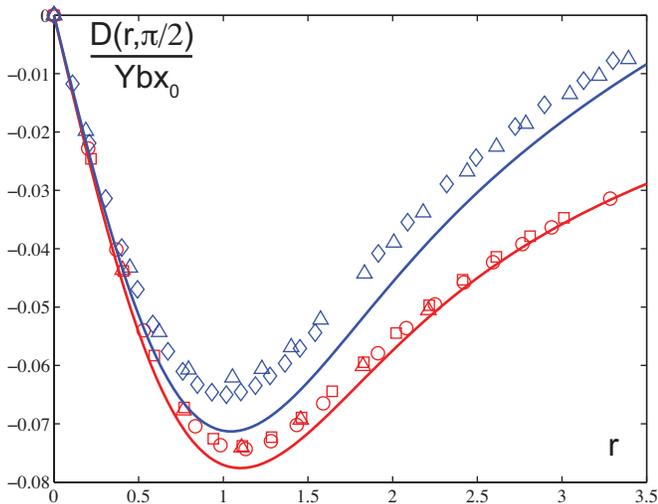}
\caption{\label{fig:dislocpotential_alpha1} The dislocation potential $D(r,\pi/2)$ in Eq.(\ref{eq:D1}) is plotted as a continuous line for a Gaussian bump parameterized by $\alpha = 1$.  Open symbols represent the numerical minimization of a fixed connectivity harmonic model for which the separation of the two 5-7 disclinations is fixed while sliding the dislocation as a whole radially with respect to the bump: $\theta=\pi/2$, $R/x_{0}=$4 (blue), 8 (red); $x_{0}/a=$ 10 ($\bigcirc$), 15 ($\Box$), 20 ($\triangle$), 30 ($\diamondsuit$). The R-dependent finite size corrections of Appendix \ref{app-boundary} are included.}
\end{figure}
In this work, we concentrate primarily on the physics of dislocations, vacancies, interstitials and impurities, which (as confirmed by our simulations) is adequately described by the formalism embodied in Equations (\ref{eq:vonKarman}) and (\ref{eq:chi}). Our analytical results are compared whenever possible to numerical minimizations of an harmonic lattice model; these computations corroborate our qualitative conclusions even beyond the limits $\alpha \ll 1$ and $a \ll x_{0}$ for which Equations (\ref{eq:vonKarman}) and (\ref{eq:chi}) are strictly valid.
For example, Fig. \ref{fig:F0} illustrates how our theoretical predictions of the frustration energy $F_{0}$, based on Eq.(\ref{eq:F0}), fit numerical data for increasingly larger values of $\alpha$.   
Similarly, rather good agreement between theoretical predictions and numerics is obtained for the dislocation potential $D(r,\theta)$ even if $\alpha=1$, as illustrated by Fig. \ref{fig:dislocpotential_alpha1}. 

\section{\label{app-boundary} Curvature induced finite size effects}
In this appendix, we discuss how the geometric potentials derived above for an infinite system are modified by the presence of a circular boundary 
of radius $R$ on which free boundary conditions apply  
\begin{equation}
n_{i} \sigma_{ij}(R)= \frac{1}{x_{0} R} \left[\frac{\partial \chi}{\partial r} \right]_{r=\frac{R}{x_0}}=0 \!\!\!\! \quad .
\label{eq:boundary-condition}
\end{equation}   
where $n_{i}$ is the unit normal to the circumference of the system. We first determine the Airy function $\chi^{G}(r)$ that describes elastic deformations caused by the Gaussian curvature $G(r)$ only, without any contributions from the defects. As explained in Section \ref{frustration}, once $\chi^{G}(r)$ is known the energy of geometric frustration can be easily calculated. 
We start by fixing the harmonic function $H_{R}(\vec{x})$ introduced in Eq.(\ref{eq:vonKarman2}) upon using the boundary condition of Eq.(\ref{eq:boundary-condition}). Note that by azimuthal symmetry, the Gaussian curvature $G(r)$ and hence $H_{R}(r)$ are constant on the circular boundary. This allows us to conclude that the harmonic function is constant
everywhere; we denote this constant by $H_{R}$. The subscript indicates that the constant $H_{R}$ is a function of the system size $R$. We can explicitly determine $H_{R}$ by integrating 
\begin{equation}
\frac{1}{Y} \Delta \chi ^{G}(r)= -V(r) + H_{R} \!\!\!\! \quad ,
\label{eq:vonKarman2-bis}
\end{equation}   
over the area of the circular disk, with the result 
\begin{equation}
\left[r \frac{\partial \chi}{\partial r} \right]^{R} _{0}=Y x_0^2 \int ^{R/x_0} _{0}  \left[H_{R} -V(r)\right] r dr \!\!\!\! \quad .
\label{eq:interm-appA}
\end{equation}   
Upon substituting Equations (\ref{eq:boundary-condition}) and (\ref{eq:V}) into Eq.(\ref{eq:interm-appA}), we obtain by integration
\begin{equation}
H_R=\frac{\alpha ^2 x_{0}^2}{4 R^2}\left[e^{-\left(\frac{R}{x_{0}}\right)^2}-1\right] \!\!\!\! \quad ,
\label{eq:HR}
\end{equation}   
which insures that the forces on the boundary vanish. Note that $\sigma^{G}_{r\phi}=\frac{1}{x_0^2}\frac{1}{r^2}\frac{\partial ^2 \chi ^G}{\partial \phi ^2}=0$ as a consequence of azimuthal symmetry, so the second boundary condition is automatically satisfied. For large system sizes $R \gg x_{0}$, $H_{R} \simeq -\frac{\alpha ^2 x_{0}^2}{4 R^2}$.

Upon using the general definition of the Airy function in the same limit, we obtain
\begin{equation}
\label{eq:sigma-bis}
\sigma^{G}_{kk}(r) = \Delta \chi^{G}(r) = -YV(r) - \frac{\alpha ^2 Y x_{0}^2}{4 R^2} \!\!\!\! \quad  .
\end{equation}
Substitution of Eq.(\ref{eq:sigma-bis}) in Eq.(\ref{eq:chi}) gives an estimate of the stretching energy, $F_{0}$, of the defect-free crystal that accounts for finite size effects, namely
\begin{equation}
   F_{0} \simeq \frac{Y}{2} \int \! dA \!\!\!\! \quad
   \left[V(r) + \frac{\alpha ^2  x_{0}^2}{4 R^2}\right]^2    \!\!\!\! \quad .
   \label{eq:F0-bis}
\end{equation}
The graph in Fig. \ref{fig:F0} is a numerical plot of $F_{0}$ versus $\alpha$ for $\frac{R}{x_{0}}$ equal to 8 and 4 that corroborates the results of Eq.(\ref{eq:F0-bis}).
This result is obtained from the one quoted in the main text by simply performing the substitution $V(r) \rightarrow V(r)-H_{R}$ in Eq.(\ref{eq:F0}).
Note that the form of $H_{R}$ was determined by solving for $\chi^{G}(r)$ in Eq.(\ref{eq:vonKarman2-bis}). Strictly speaking, in the presence of defects one should
solve the full Eq.(\ref{eq:vonKarman}) that accounts for the presence of defects by 
means of an extra source term. The solution $\chi (\vec{x})$ would not be azimuthally symmetric especially when defects are located well away from the center of the bump. 
However, as detailed in Sections \ref{sec:dislocpotential} and \ref{sec:vacancies}, the geometric forces 
experienced by the defects at different locations on the bump can be easily calculated once $\sigma^{G}_{ij}(r)$ is known without solving 
the full biharmonic equation provided that $x_{0} \ll R$. The leading finite size correction to $\sigma^{G}_{ij}(r)$ was indeed calculated
in Eq.(\ref{eq:sigma-bis}).

We can thus make progress in the calculation of the defect potentials in the presence of the boundary by simply letting $V(r) \rightarrow V(r)-H_{R}$ in 
Eq.(\ref{eq:intermediate-step}). The defect potential now reads
\begin{eqnarray}
\zeta (\vec{x}^{\alpha}) = -Y \int \! dA' S(\vec{x}') \!\!\!\! \quad \int \! dA \frac{1}{\Delta_{\vec{x}\vec{x}'}} \!\!\!\! \quad \left[ V(\vec{x})-H_{R} \right] \!\!\!\! \quad . 
\label{eq:intermediate-step-bis}
\end{eqnarray} 
The result for the geometric potential
of dislocations $D(r,\theta)$ with finite size effects (see Section \ref{sec:dislocpotential}) becomes   
\begin{eqnarray}
   D(r,\theta) &=& -Y b_{i} \!\!\!\!\! \quad \epsilon_{ij} \!\!\!\!\! \quad \partial_{j} \int \! dA' \!\!\!\! \quad \frac{1}{\Delta_{\vec{x}\vec{x'}}} \!\!\!\! \quad
  \left(V(r')+\frac{\alpha ^2 x_{0}^2}{4 R^2}\right)  \nonumber\\ & \approx & Y \!\!\!\!\! \quad  b \!\!\!\!\! \quad x_{0} \!\!\!\!\! \quad \frac      
         {\alpha^{2}}{8}  \sin{\theta} \left[ \left( \frac{e^{-r^{2}} - 1}{r}\right) + \left(\frac{x_{0}}{R}\right)^2 r \right] \!\!\!\! \quad . \label{eq:D1-bis}
\end{eqnarray}
The second term in $D(r,\theta)$ can be viewed as the geometric potential of an image dislocation in an infinite system (let $R \rightarrow \infty$ in Eq.(\ref{eq:D1})) evaluated at the image position of $|\vec{x}|'=\frac{R^2}{|\vec{x}|}$ and with Burger's vector $\vec{b'}=\vec{-b}$. Thus, the dislocation interacts with the curvature directly and via its image. This choice of an image defect insures that the stress normal to the boundary, $\sigma_{rr}=\frac{1}{x_0^2}\frac{1}{r} \frac{\partial \chi}{\partial r}$, vanishes.  
The curvature induced boundary term included in 
Eq.(\ref{eq:D1-bis}) has no analogue in flat space but its effect is evident upon examining the results of our curved space simulations presented in Fig. \ref{fig:dislocpotential} for two different choices of $\frac{R}{x_{0}}$ equal to 8 and 4 and aspect ratio $\alpha=0.5$. The comparison between numerics and the analytical expression derived in Eq.(\ref{eq:D1-bis}) confirms the validity of our approximation scheme.

\begin{figure}
\includegraphics{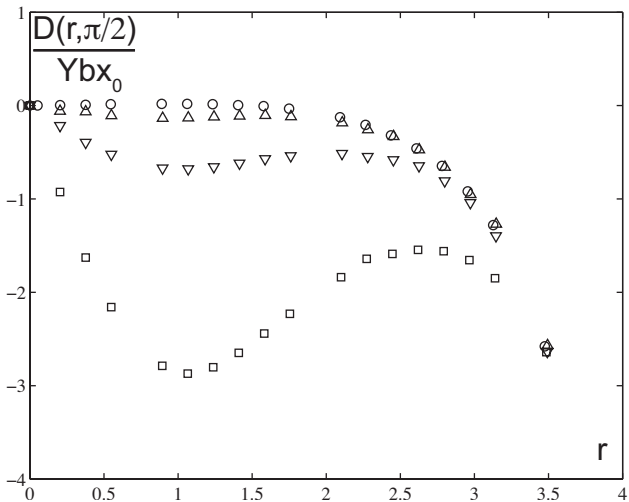}
\caption{\label{fig:alphaS} 
The dislocation potential $D(r,\pi/2)$ is plotted versus the scaled coordinate $r = x/x_{0}$, $x_{0} = 10 a $, $R=40 a$ for $\alpha = $ 0 ($\bigcirc$), $0.05$ ($\triangle$), $0.1$ ($\nabla$) and $0.2$ ($\Box$).  The numerical energy is scaled by $10^{-3} x_{0}$, and shifted so that $D(0)\equiv0$.  Notice that the minimum is gradually lost for $\alpha < \alpha_S \approx 0.1$.
}
\end{figure}
We should stress that the image defect does $not$ guarantee that the second boundary condition $\sigma_{r \phi}=0$ is satisfied when the presence of an off-axis dislocation breaks the azimuthal symmetry. However, we argue qualitatively that, as the dislocation approaches the boundary, it will be attracted to it as an electrostatic charge would to its image.  A systematic treatment of this boundary induced self energy $B(r,\theta)$ for arbitrary orientations of the Burger vector of the dislocation is outside of the scope of this work. Here we will only provide some estimates of how $B(r,\theta)$ (present when $R$ is finite) wins out over the geometric interaction $D(r,\theta)$ in the flat space limit $\alpha \rightarrow 0$. The geometric potential that tries to confine the 
dislocation close to the bump, scales like $D(r,\theta) \sim Y b x_{0}$ (see Eq.(\ref{eq:D1-bis})), whereas the attraction to the ``image dislocation" is approximately given by $B(r,\theta) \approx  Yb^2 \left(\frac{|\vec{x}|}{R} \right)^2$. This choice insures that the boundary interaction is unchanged if the position of the defects is flipped $\vec{x}\rightarrow - \vec{x}$ and suggests that $B(r,\theta)$ evaluated at the minimum of $D(r,\theta)$ is given approximately by $\frac{Y b^2 x_{0} ^2}{R^2}$ (to leading order in a perturbation expansion in the small parameter $\frac{x_{0}}{R}$). A simple force (or equivalently an energy balance) argument
provides an estimate of the spinoidal aspect ratio $\alpha_{s} \sim \frac{\sqrt{b x_{0}}}{R}$ below which the minimum of the geometric 
potential $D(r,\theta)$ is lost to the boundary attraction $B(r,\theta)$. Our estimate agrees with the numerical results
illustrated in Fig. \ref{fig:alphaS}. 

\section{\label{app-numerics} Numerical Methods}

We have complemented our analytical studies with numerical minimizations of the  in-plane stretching energy of a triangular lattice of points connected by springs draped over the Gaussian bump described in the text.  The discrete stretching energy for a set of points at positions $\vec{r}_{\mu}$ is defined by
\begin{equation}
   \label{eq:Fdis}
   F^{\mathrm discrete} = \frac{1}{4} \sum_{\mu,\nu} k_{\mu,\nu} (r_{\mu,\nu} - a)^2,
\end{equation}
where $r_{\mu,\nu} = |\vec{r}_{\mu} - \vec{r}_{\nu}|$, $\vec{r}_{\mu} = (x_{\mu},y_{\mu},h(x_{\mu},y_{\mu}))$, and $a$ is the lattice spacing.  The height function $h(x,y)$ defines the fixed topography of the system and is given by $h(x,y) = \alpha x_{0} e^{-(x^2 + y^2)/2x_{0}^2}$.  The spring constant matrix is $k_{\mu,\nu} = k n_{\mu,\nu}$, where the connectivity matrix $n_{\mu,\nu}$ specifies that the underlying lattice is triangular, with a fully coordinated particle having 6 nearest neighbors (n.n.)
\begin{equation}
  n_{\mu,\nu} = \left\{ \begin{array}{l} 1 \quad \mathrm{if} \quad \mu,\nu \quad \mathrm{n.n.} \\ 0 \quad \mathrm{otherwise} \end{array} \right. .
\end{equation}
Defects are introduced into the lattice by changing the number of nearest neighbors from 6 to 5/7 for +/- disclinations, respectively.  Dislocations, interstitials and vacancies are composed of specific configurations of +/- disclinations.  By coarse-graining Eq.(\ref{eq:Fdis}) \cite{seun88}, we can make an explicit connection to the macroscopic energy formula (\ref{eq:lame}) with Young modulus, $Y = \frac{2}{\sqrt{3}} k $ and Poisson ratio, $\sigma = 1/3$, from which the Lam\'{e} coefficients $\lambda$ and $\mu$ can be determined \cite{Landau-book}.  In this work, we use units of length and energy such that $a = 1$ and $k = 1$.

\begin{table*}[ht]
\caption{\label{tab:partConstr} Particles constrained in energy minimizations.}
\begin{tabular}{ccc}
\hline
Configuration & Number of Particles Fixed & Particles Fixed \\
\hline
Defect Free & 1 & Particle at (x,y) = (0,0) \\
Isolated Dislocation & 2 & +/- Disclination pair \\
Two Opposing Dislocations & 4 & +/- Disclination pairs \\
Interstitials/Vacancies (I/V) & 1 & Particle at (x,y) = (0,0) \\
\hline
\end{tabular}
\end{table*}

Numerical minimizations of Eq.(\ref{eq:Fdis}) are performed for circular patches of triangular lattice draped over a bump, with frozen-in defect configurations.  Thus every point in Figure \ref{fig:dislocpotential}, for example, represents a separate energy minimization.  To perform the minimization, we use the standard Fletcher-Reeves (FR) conjugate-gradient method \cite{NumRecip,FRBook} in which particles are moved along successive directions determined by the gradient of the energy in Eq.(\ref{eq:Fdis}) with respect to particle coordinates. \footnote{The FR algorithm requires the use of a one-dimensional minimization algorithm, for which we used the Brent algorithm \cite{NumRecip} as implemented by the Gnu Scientific Library \cite{GSL}.  The Brent algorithm requires bounds to be placed on the dimensionless parameter $\lambda$, which we typically take to be $-5 < \lambda < 10$.}
After taking into account the fact that the $z$-coordinate is determined by $h(x,y)$, the gradient with respect to the $x$-coordinate of particle $\eta$ reads
\begin{widetext}
\begin{equation}
   \label{eq:FdisGrad}
   \frac{\partial F}{\partial x_{\eta}} = \sum_{\mu,\nu} \frac{k_{\mu,\nu} }{2} 
      \frac{(r_{\mu\nu} - a)}{r_{\mu\nu}} (\delta_{\mu,\eta} - \delta_{\nu,\eta})
      \left\{ (x_{\mu} - x_{\nu}) + \left[ h(x_{\mu},y_{\mu}) - h(x_{\nu},y_{\nu}) \right]  
           \frac{\partial}{\partial x_{\eta}} h(x_{\eta},y_{\eta}) \right\},
\end{equation}
\end{widetext}
with $\delta_{\mu,\nu}$ the Kronecker delta. To obtain the gradient with respect to the y-coordinate, interchange $x \leftrightarrow y$.

Convergence in the gradient is achieved  when the magnitude of the gradient of the energy drops below some defined tolerance, which was set to $10^{-5}$ ($k=1$).  In order to ensure a smooth approach to the energy minimization, gradient convergence was accepted only if the $|\Delta F^{\mathrm{discrete}}|$ between the last two iterations of the algorithm was less than $10^{-8}$ ($k=1$). 

Since there is a non-zero frustration energy for a defect-free lattice on a bump, the particles will collectively slide off of the bump into flatland if allowed.  To prevent this, some particles were always fixed during minimization as indicated in Table \ref{tab:partConstr}.  These constraints were implemented so that a particle was always located at the center of the bump.

\end{document}